\documentclass[epj]{svjour}
\usepackage{graphics}
\begin{document}
\title{Hollywood blockbusters and long-tailed distributions}
\subtitle{An empirical study of the popularity of movies}
\author{Sitabhra Sinha\inst{1}%
\thanks{e-mail: sitabhra@imsc.res.in}%
\and S. Raghavendra\inst{2}
\thanks{e-mail: raghav@mse.ac.in}%
}                    
\institute{Institute of Mathematical Sciences, C. I. T. Campus,
Taramani, Chennai - 600 113, India. \and Madras School of Economics,
Anna University Campus, Chennai - 600 025, India.}
\abstract{
Numerical data for all movies released in theaters in the USA during 
the period 1997-2003
are examined for the distribution of their popularity in terms of 
(i) the number of weeks they spent in the Top 60 according to the weekend 
earnings, and (ii) the box-office gross during the opening week, as
well as, the total duration for which they were shown in theaters.
These distributions show long tails where the most popular movies are located.
Like the study of Redner [S. Redner, Eur. Phys. J. B \textbf{4}, 131 (1998)]
on the distribution of citations to individual papers, 
our results are consistent with
a power-law dependence of the rank distribution of gross revenues for 
the most popular movies with a exponent close to $-1/2$.
\PACS{
      {89.75.Da}{Systems obeying scaling laws}   \and
      {89.65.-s}{Social and economic systems}   \and
      {02.50.-r}{Probability theory, stochastic processes, and statistics}
     } 
} 
\maketitle
In recent times there has been a surge of interest in applying statistical
mechanics to understand socio-economic phenomena \cite{Dur99}. 
The aim is to seek
out patterns in the aggregate behavior of interacting agents, which
can be individuals, groups, companies or nations. Examples
of such patterns arising in a social or economic context include the
Pareto law of income distribution \cite{pareto}, 
Zipf's law in the distribution of firm sizes \cite{Sta95}, etc.
Another fruitful area for seeking such patterns is the evolution of
collective choice from individual behavior, e.g., the sudden emergence of
popular fads or fashions \cite{Bik92}. The popularity or `success' of
certain ideas or products, compared to
their numerous (often very similar) competitors, cannot be explained
exclusively on the basis of their individual merit. 
Empirical investigation of such popularity distributions may shed
light on this issue. In particular, they
can be used to test different theories of
how collective choice emerges from individual decisions based on
limited information and communication among agents \cite{Arrow}.
With this objective, we have investigated in this paper
the popularity of movies by
estimating the distributions of their gross earnings (opening and total)
and their endurance in the box office. 
Our results are consistent with
a power-law dependence of the rank distribution of gross revenues for 
the most popular movies, with an exponent close to $-1/2$.

A number of recent papers have looked at the empirical distribution of
popularity or `success' in different areas. Redner \cite{Red98} has 
analyzed the distribution of citations of individual papers and has
found that the number of papers with $x$ citations, $N ( x )$ has
a power law tail $N ( x ) \sim x^{-3}$. This is consistent with his
observation that the Zipf plot of the number of
citations against rank has a power law dependence with exponent $\sim -1/2$.
In contrast, Laherr\`ere and Sornette \cite{Lah98}
have looked at the lifetime total citations
of the 1120 most cited physicists, and Davies \cite{Dav02}
at the lifetime total
success of popular music bands as measured by the total number of weeks
they were in the weekly top 75 list of best-selling recordings. Both report
the occurrence of stretched exponential distribution. Teslyuk $et al$
\cite{Tes04} have focussed on the popularity of websites, and have
described the rank distribution by a modified Zipf law.
In the specific context of movie popularity, De Vany and Walls have looked at
the distribution of movie earnings and profit as a function of a variety
of variables, such as, genre, ratings, presence of stars, etc. \cite{DeV03}.
They have shown that the distribution of box-office revenues follow a
Levy stable distribution \cite{DeV99} arising from Bose-Einstein dynamics
in the information feedback among movie audiences \cite{DeV96}. Stauffer and
Weisbuch \cite{Sta03} have tried to reproduce the observed rank distribution
of top 250 movies (according to votes in www.imdb.com) using a social 
percolation model.

For our analysis we decided to look at all movies released in theaters
in the United States during the period 1997-2003. 
These include not only new movies produced in the USA in this period,
but also re-release of older movies as well as movies made abroad \cite{note}.
However, perhaps unsurprisingly, the top performing movies (in terms of 
box-office earnings) almost invariably are products of the major
Hollywood studios.
The primary database
that we used was The Movie Times website \cite{movietimes} which
listed the movies released during these years and, for the period 1999-2003,
had information concerning the opening 
and total gross and the number of weeks the movie stayed
at Top 60 according to the weekend earnings. The corresponding
data for 1997-98 was obtained from the Internet Movie 
Database \cite{imdb}. Table \ref{table1} gives all the relevant details
concerning the data set used for the following analysis.

\begin{figure}
\resizebox{0.5\textwidth}{!}{%
  \includegraphics{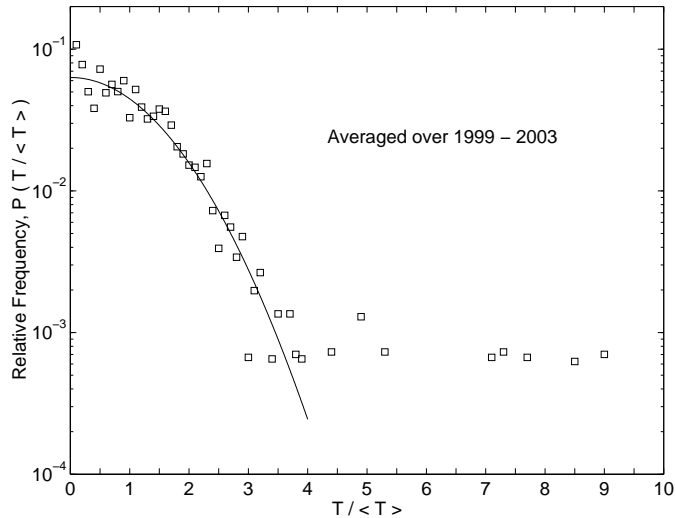}
}
\caption{Normalized relative frequency distribution of 
number of weeks in Top 60
divided by the average number of weeks spent by movies in Top 60
in a year. The frequency distribution is computed for each year in
the period 1999-2003 and then averaged over the entire period. 
The curve represents a gaussian distribution fitted over the data.}
\label{fig1}       
\end{figure}

As a first measure of popularity we look at the number of weeks a movie 
spent in the Top 60. While this quantity may superficially seem similar
to that observed by Davies for popular musicians \cite{Dav02}, note that
we are looking at the popularity of individual products (releases) and
not the overall popularity of the producer (performer).
Figure \ref{fig1} shows the relative frequency distribution of 
the number of weeks a movie spent in Top 60, scaled by its average for a 
given year, and then averaged over the period 1999-2003. The period
of one year was chosen to remove all
seasonal variations in moviehouse attendance, e.g., the peak around
Christmas. The data for less popular movies could be fitted very well
with a normal distribution. However, the more popular movies reside
at the long tail of the distribution and cannot be explained by a 
gaussian process. 

\begin{figure}
\resizebox{0.5\textwidth}{!}{%
  \includegraphics{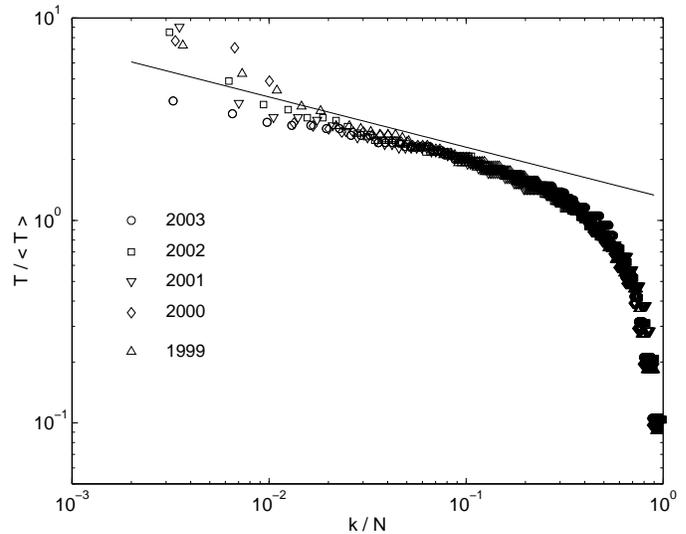}
}
\caption{Zipf plot of the number of weeks, $T$, spent in Top 60 by the
$k$-th ranked movie for the years 1999-2003.
The rank $k$ has been scaled by $N$, the total number of movies
released in theaters
that year, while $T$ has been scaled by its annual average.
A straight
line of slope $- 0.248$ is shown for visual reference.}
\label{fig2}       
\end{figure}

\begin{table}
\caption{Annual data for movies released across theaters in USA
for the period 1997-2003:
the 2nd column represents the number of movies released in the year, $N$;
the 3rd column is the average number of weeks a movie spent in Top 60
(in terms of weekend gross); the 4th and 5th columns represent the
average opening and total gross, respectively, for movies released
in a particular year. The general trend, with a few exceptions, seems to be 
that both opening and total
gross averages increase with time. (N.A. = not available)}
\label{table1}       
\begin{center}
\begin{tabular}{|c|c|c|c|c|}
\hline
 & & & & \\
~~Year~~ & ~$N$~ & ~$< T >$~ & ~$< G_O >$~ & ~$< G_T >$~  \\
 & & (weeks) & (in M\$) & (in M\$)\\
\noalign{\smallskip}\hline\noalign{\smallskip}
2003 & 307 & 9.5 & 8.094 & 29.239\\
2002 & 320 & 9.6 & 7.468 & 28.440\\
2001 & 285 & 10.5 & 7.332 & 28.331\\
2000 & 299 & 10.2 & 6.155 & 25.470\\
1999 & 274 & 10.9 & 5.638 & 26.452\\
1998 & 260 & N.A. & 6.389 & 23.951\\
1997 & 289 & N.A. & 5.735 & 26.108\\
\noalign{\smallskip}\hline
\end{tabular}
\end{center}
\end{table}
The scarcity of data points in the tail meant that one could not
infer the exact dependence from the relative frequency distribution alone.
We, therefore, looked at the rank ordering statistics which focuses
on the largest members of the distribution 
(the most popular movie being ranked 1). 
As has been noted
previously, the exponent of a power-law distribution can be determined
with good accuracy in such a plot, even with relatively few data points
\cite{Red98,Lah98}.
Fig. \ref{fig2} shows a rank ordered plot of the scaled time that a movie
spent in the Top 60. The ranks ($k$) have been scaled by the total number of 
movies ($N$) that were released in a particular year. Note that the data for 
all the years 1999-2003 appear to follow the same curve (excepting
for the top ranked movies). A power-law distribution fitted to this
data gave an exponent of $\simeq -0.248$. The result implies that
while the endurance of less popular movies seems to be a stochastic
process, the longevity of more popular movies at the box office
is possibly due to interactions among agents (moviegoers)
through a process of information transfer. This could be
responsible for the deviation from a gaussian distribution and the
formation of a long tail following a power law.

However, a movie residing in the Top 60 for a long time  does not 
necessarily imply that it was
seen by a large number of people. A few of the longest running movies
were films designed for specialized projection theaters having giant
screens, e.g., in our dataset the movie which spent the maximum time 
in the Top 60 (95 weeks) was ``Shackleton's Antarctic Adventure''
that was being shown at Imax theaters. In terms of gross earnings,
these movies performed poorly. Therefore, we decided to look at
the box-office revenues of movies, both for the opening week and 
for the total duration it was shown at theaters. 
Although total gross may be a better measure of movie popularity,
the opening gross is often thought to signal the success of a 
particular movie. This is supported by the observation that about
65-70 $\%$ of all movies earn their maximum box-office revenue in the
first week of release \cite{DeV99}.

\begin{figure}
\resizebox{0.5\textwidth}{!}{%
  \includegraphics{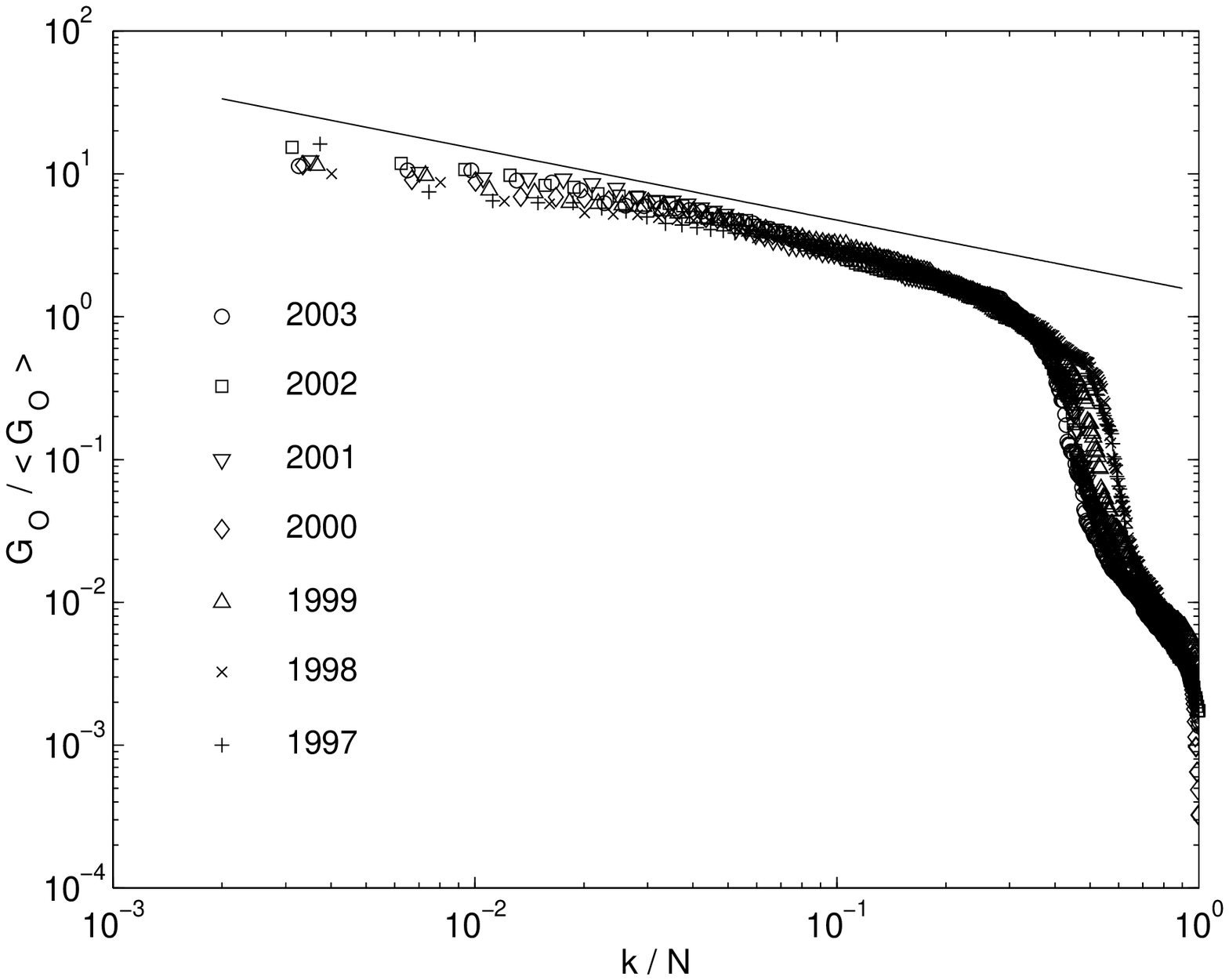}
}
\resizebox{0.5\textwidth}{!}{%
  \includegraphics{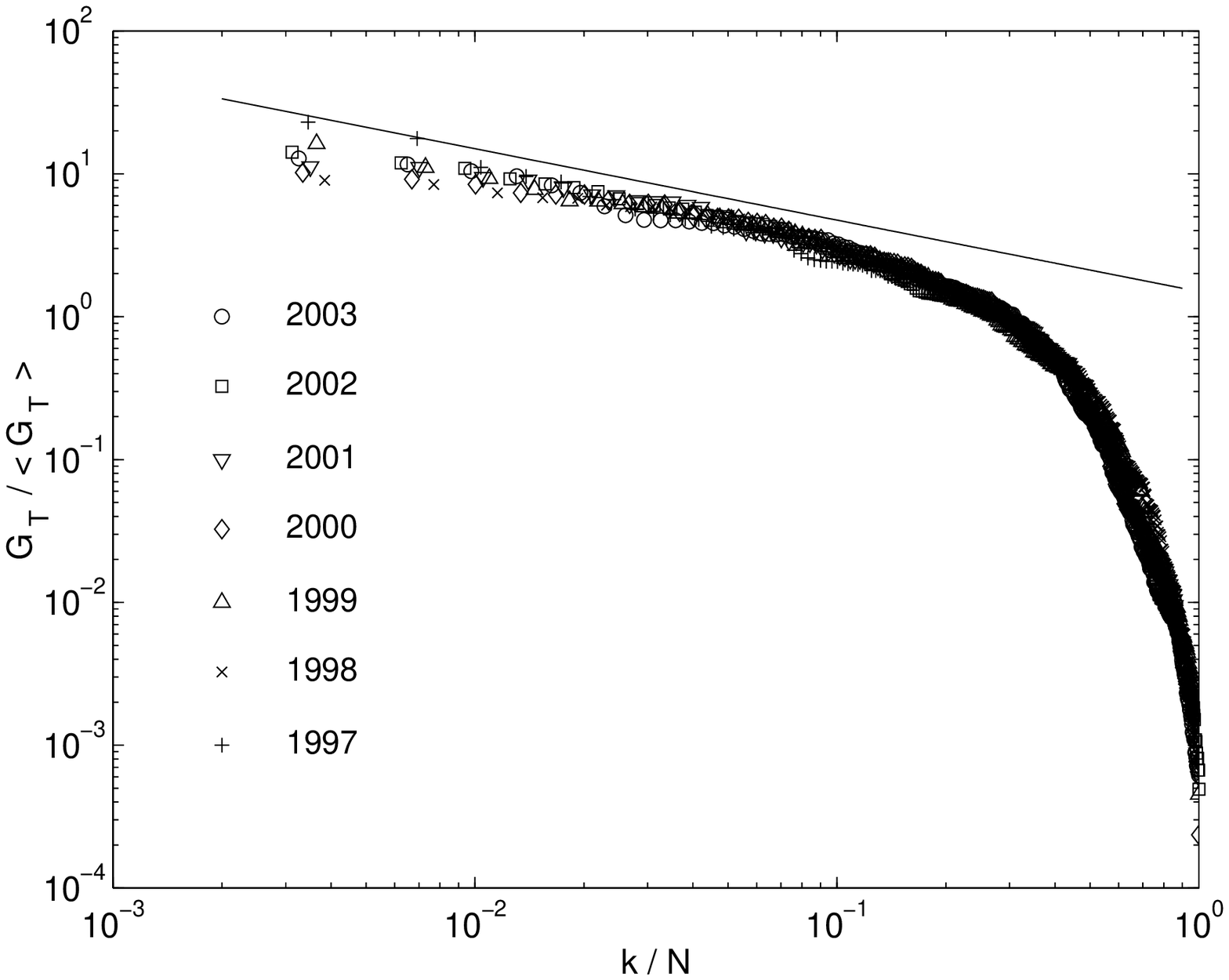}
}
\caption{Zipf plots of the scaled rank distribution of movies according
to the opening gross (top) and total gross (bottom) for the years 1997-2003. 
The rank $k$ has been scaled by the total number of movies released that
year ($N$), 
while the gross ($G_O$, $G_T$) has been scaled by its annual average.
Straight
lines of slope $- 0.5$ are shown for visual reference.}
\label{fig3}       
\end{figure}

To correct for inflation, we scaled the gross earnings by the average values
for a particular year. 
The relative frequency distributions had too
few points at their extremities for a reasonable determination
of the nature of the tails. 
For better resolution of the distribution at
the tails, we looked at the Zipf plots (Fig. \ref{fig3}).
Scaling the rank ($k$) by the total number
of movies released ($N$), and the gross by its average for that year,
led to the data for all years collapsing onto the same curve. This 
indicates that the distribution is fairly stable across the period 
under study. 
The data for the opening, as well as the total gross, show a power
law distribution with an exponent $\sim - 1/2$ in the region where
the top grossing movies are located. 

The only difference between the opening and the total
gross Zipf plots occur at the region of poorly performing movies, with
a kink in the former that indicates the presence 
of bimodality in the opening gross distribution \cite{Sin04}.
Based on this we conclude that, movies in their opening week, either
perform very well, or very poorly. However, some movies, though not
popular initially, may generate interest over time and 
eventually become successful in terms of total revenue earned. In
movie parlance, these are known as ``sleeper hits''. This can be seen
from the total gross distribution becoming unimodal, showing
a smoother curvature than the opening gross distribution in the Zipf plot.

To verify whether the data is better explained by a stretched 
exponential distribution, we have fitted the cumulative relative 
frequency distribution
of scaled total gross, $G_T / <G_T>$, to a function of the form
$P_c ( x ) \sim$ exp$[-(x/x_0)^\beta]$, with $x_0 = 1$ and $\beta = 0.67$
for best fit. However, the rank distribution curve obtained for
these parameter values did not describe well the corresponding
empirical data over the entire range. A similar exercise was carried
out for the opening gross data which gave different parameter values
for best fit.
As in the case of total gross, these also failed in describing 
the opening gross rank distribution over the entire range.

\begin{figure}
\resizebox{0.5\textwidth}{!}{%
  \includegraphics{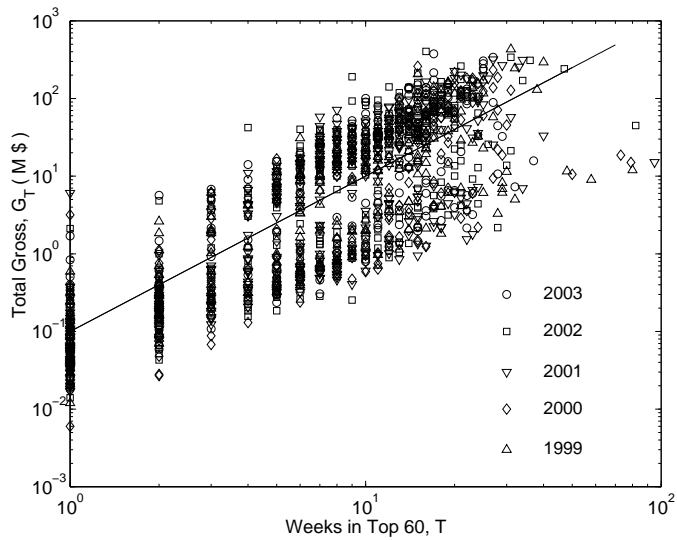}
}
\caption{Plot of the
number of weeks spent in Top 60 against the total gross earned
by movies released during the years 1999-2003. A straight line of
slope 2 is shown for visual reference. Note that the few outliers
on the right of the figure (with large values of $T$) correspond almost
exclusively to movies specially produced for screening in Imax theaters.}
\label{fig4}       
\end{figure}

The occurrence of different exponent values for the distribution of
time spent in Top 60 and the gross distributions may initially seem 
confusing. To resolve this issue we looked at the total gross of a movie,
$G_T$, against the number of weeks that it spent in the Top 60, $T$ 
(Fig. \ref{fig4}).
All movies released during 1999-2003 were used to generate the figure.
Plotting on log-log scale yielded a relationship that implied
$G_T \sim T^{2.087}$, which is consistent with the exponent obtained from
gross distribution being approximately
twice that of the exponent obtained from
the distribution of number of weeks spent in Top 60.

%
We have also looked at the distribution of movie popularity according to 
the number of votes they received from registered users of IMDB \cite{imdb}. 
The Zipf plot
of the votes against the movie rankings for the top 250 movies as of
May 9, 2004, did not seem to follow a single functional relation over the
full range. However, the middle range seemed to fit an exponential
distribution. Note that this popularity measure is very different
from the ones we have used above, as in this case, most of the
movies in the top 250 list are very well-known and a large amount 
of information is available about them. On the other hand, the
movies that have been released recently are relatively unknown and
people often make their decisions to watch them on the basis of 
incomplete and unreliable information.

We want to point out that the gross distributions of individual films
is similar in nature to the citation distribution of scientific papers
investigated by Redner \cite{Red98}. It is of interest to note that 
he also obtained an exponent of $-1/2$, in the very different context
of a Zipf plot of the number of citations to a given paper
against its citation rank. This may be indicative of an universal
feature, as both these cases are looking at how success or 
popularity is distributed
in different areas of human creativity. In both cases, an individual
entity (paper or movie) becomes popular, or successful, as a result
of information propagation in a community. The influence of this
information on individual choice, and the resulting actions 
of a large number of individuals, leads to the collective response
of the community to the entity. To be popular, an entity needs to 
generate a large
number of favorable responses. Clearly, while most such entities 
elicit a stochastically distributed number of favorable responses, 
a few manage to generate enough initial popularity which then
gets amplified through interactions among agents
to make it even more popular. 
In other words, the interactions cause the distribution to deviate from 
that of a purely random process, resulting in the long tails seen in
the popularity distributions.

We would like to thank D. Stauffer for arousing our interest in this topic
and B. K. Chakrabarti for critical comments.

\end{document}